\documentclass[aps,preprint]{revtex4}

\usepackage{epsfig,amsmath}
\usepackage{subfigure}
\usepackage{graphicx}
\usepackage{dcolumn}
\usepackage{stmaryrd}
\usepackage{mathrsfs}
\usepackage{pifont}
\usepackage{amsthm}
\usepackage{amssymb}
\usepackage{bm}
\usepackage{latexsym}
\usepackage{hyperref}
\usepackage{color}

\begin{document}

\title{Determine Ramsey numbers on a quantum computer}
\author{Hefeng Wang}
\thanks{Correspondence to wanghf@mail.xjtu.edu.cn}
\affiliation{Department of Applied Physics, Xi'an Jiaotong University, Xi'an 710049, China}

\begin{abstract}
We present a quantum algorithm for computing the Ramsey numbers whose
computational complexity grows super-exponentially with the number of
vertices of a graph on a classical computer. The problem is mapped to a
decision problem on a quantum computer, a probe qubit is coupled to a
register that represents the problem and detects the energy levels of the
problem Hamiltonian. The decision problem is solved by determining whether
the probe qubit exhibits resonance dynamics. The algorithm shows a quadratic
speedup over its classical counterparts, and the degenerate ground state
problem in the adiabatic quantum evolution algorithm for this problem is
avoided.
\end{abstract}

\maketitle

\emph{Introduction.}--Ramsey numbers are part of the Ramsey theory~\cite%
{ramsey} which is a branch of mathematics that studies the occurrence of
order in large disordered structures, and has a wide application in
mathematics, information theory, and theoretical computer science~\cite%
{rosta,paris,fur}. The computation of Ramsey numbers are extremely
difficult, however, only nine of them are currently known. Here, we focus on
the two-color Ramsey numbers which can be described as follows~\cite{ramsey}%
: in an $n$-vertex graph, the $x$ vertices form an $x$-clique, and the $y$
vertices form an $y$-independent set. An $x$-clique is a set of $x$ vertices
in which any two of the vertices are connected by an edge, while an $y$%
-independent set is a set of $y$ vertices in which no two of the vertices
are connected by an edge~\cite{boll}. According to the Ramsey theory~\cite%
{ramsey}, there exists a threshold value $R(x,y)$ for given integers $x$ and
$y$, such that every graph of $n$ vertices contains either an $x$-clique, or
an $y $-independent set as long as $n\geq R(x,y)$. The task is to compute
the Ramsey number--the threshold value $R(x,y)$ for given $x$ and $y$.

A total number of $2^{n(n-1)/2}$ different graphs can be formed by $n$
vertices. To determine whether $n$ is the Ramsey number $R(x,y)$, one has to
check all the $2^{n(n-1)/2}$ graphs, which grows super-exponentially with $n$%
, and the task quickly becomes intractable. Mathematically, bounds for
Ramsey numbers have been given, but it is still a challenge to determine the
exact Ramsey numbers in most cases.

In Ref.~\cite{gaitan}, the problem of computing a Ramsey number is mapped to
a combinatorial optimization problem, and an adiabatic quantum
optimization~(AQO) algorithm is applied for finding the solution. The
algorithm has been implemented experimentally on the D-Wave One device~\cite%
{bian}. In this algorithm, a Hamiltonian for the problem is constructed, and
the system is evolved to the ground state of the problem Hamiltonian through
adiabatic quantum evolution. The task is transformed to check whether the
ground state energy of the problem Hamiltonian equals or larger than zero.
In this algorithm, one has to obtain the ground state of the problem
Hamiltonian first then calculate the ground state energy. For a Ramsey
number, it usually corresponds to a number of graphs, which means the ground
state of the problem Hamiltonian are usually degenerate. The runtime of the
AQO algorithm is determined by the minimum energy gap between the ground
state and the first excited state~\cite{farhi}, and it is an open question
in adiabatic quantum computing to determine how the runtime scales when the
energy gap is zero~\cite{gaitan}.

In this paper, we propose a different quantum algorithm for computing Ramsey
numbers and obtaining all the corresponding graphs that have the minimum
number of $x$-cliques or $y$-independent sets. In this algorithm, a Ramsey
number can be determined without knowing the ground state of the
Hamiltonian, therefore the degenerate ground state problem in AQO is avoided.

\emph{The algorithm.}--We use the procedure introduced in Ref.~\cite{gaitan}
to transform the problem of computing a Ramsey number $R(x,y)$ to a decision
problem on a quantum computer.

In an $n$-vertex graph $G$, there are $L=n(n-1)/2$ ways of choosing a pair
of vertices $(v,v^{\prime })$, a bit variable $a_{v,v^{\prime }}$ is
associated for each pair of $(v,v^{\prime })$, and $a_{v,v^{\prime }}=1$ if $%
v$ and $v^{\prime }$ are connected with an edge and $a_{v,v^{\prime }}=0$
otherwise. Then a bit vector $\mathbf{a}=(a_{1,2},\ldots
,a_{1,n},a_{2,3},\ldots ,a_{2,n},\ldots ,a_{n-1,n})$ of length $L$ uniquely
represents an $n$-vertex graph $G$, and there are $N=2^{L}$ vectors in
total. For given integers $n$, $x$, and $y$, count the number of $x$-cliques
$C_{x}^{n}(\mathbf{a})$ and $y$-independent sets $I_{y}^{n}(\mathbf{a})$ in
an $n$-vertex graph $G$ represented by vector $\mathbf{a}$, and define an
\textquotedblleft energy\textquotedblright\ function $h_{x,y}^{n}(\mathbf{a}%
)=C_{x}^{n}(\mathbf{a})+I_{y}^{n}(\mathbf{a})$. If $n<R(x,y)$, the minimum
of the function $h_{x,y}^{n}(\mathbf{a})$ is zero, and if $n\geq R(x,y)$,
Ramsey theory guarantees that $h_{x,y}^{n}(\mathbf{a}_{0})>0$.

On a quantum computer, each bit variable $a_{v,v^{\prime }}$ is represented
by a qubit. $|z_{k}\rangle $ with $z_{k}=0$ or $1$ represents the $k$-th
bit, and all the $N=2^{L}$\ vectors are represented by the $L$-qubit vectors
$|\psi \rangle =|z_{1}z_{2}\cdots z_{L}\rangle $. These vectors form a
complete computational basis of the $L$ qubits. The problem Hamiltonian $%
H_{P}$ is defined as
\begin{equation}
H_{P}|z_{1}z_{2}\cdots z_{L}\rangle =h_{x,y}^{n}(\mathbf{a}%
)|z_{1}z_{2}\cdots z_{L}\rangle ,
\end{equation}%
$H_{P}|\psi \rangle =0$, if and only if the bit string $z_{1}z_{2}\cdots
z_{L}$ does not contain either $x$-cliques or $y$-independent sets.

Ramsey numbers are integers in a bounded range. The computation of Ramsey
numbers begins by setting $n$ equal to a lower bound for $R(x,y)$ which can
be found in a table of two-color Ramsey numbers~\cite{boll}. Increase $n$ by
one each time, check if the ground state energy of $H_{P}$ is zero or not.
The first integer $n$ for which the ground state energy $E_{1}$ of $H_{P}$
is larger than zero is the Ramsey number $R(x,y)$ for given $x$ and $y$.

From the analysis above, we can see that computation of a Ramsey number is
transformed to a problem of determining whether the ground state energy $%
E_{1}$ of $H_{P}$ is zero or not. We proposed the following quantum
algorithm for this problem.

First, we construct a quantum register $Q$ which contains one ancilla qubit
and an $L$-qubit quantum register that represents the state space of the
problem Hamiltonian of dimension $N$. A probe qubit is coupled to $Q$ and
the Hamiltonian of the entire $\left( L+2\right) $-qubit system is
constructed as%
\begin{equation}
H=-\frac{1}{2}\omega \sigma _{z}+I_{2}\otimes H_{Q}+c\sigma _{x}\otimes A,
\end{equation}%
where $I_{2}$ is the two-dimensional identity operator, $\sigma _{x}$ and $%
\sigma _{z}$ are the Pauli matrices. The first term in the above equation is
the Hamiltonian of the probe qubit, the second term is the Hamiltonian of
the register $Q$, and the third term describes the interaction between the
probe qubit and $Q$. $\omega $ is the frequency of the probe qubit~($\hbar
=1 $) and $c$ is the coupling coefficient and $c\ll \omega $. The
Hamiltonian of register $Q$ is in the form
\begin{equation}
H_{Q}=|0\rangle \langle 0|\otimes \left[ \varepsilon _{0}\left( |0\rangle
\langle 0|\right) ^{\otimes L}\right] +|1\rangle \langle 1|\otimes H_{P},
\end{equation}%
where $\varepsilon _{0}$ is a parameter that is set as a reference point to
the ground state energy $E_{1}$ of $H_{P}$. The register $Q$ is prepared in
a reference state $|\Phi \rangle =|0\rangle ^{\otimes \left( L+1\right) }$,
which is an eigenstate of $H_{Q}$\ with eigenvalue $\varepsilon _{0}$. The
operator $A=\sigma _{x}\otimes H_{d}^{\otimes L}$, where $H_{d}$ is the
Hadamard matrices. It acts on the reference state $|\Phi \rangle $ and
generates an unified superposition of the basis states.

Suppose the problem Hamiltonian $H_{P}$ has $r$ energy levels and the $i$-th
energy level is $m_{i}$-fold degenerate, then in basis of \{$|\Psi
_{0}\rangle =|1\rangle |0\rangle |0\rangle ^{\otimes n}$, $|\Psi _{i}\rangle
=|0\rangle |1\rangle |\varphi _{i}\rangle $, $i=1,2,\cdots ,r$\}, where $%
|\varphi _{i}\rangle =\frac{1}{\sqrt{m_{i}}}%
\sum_{s_{i}=0}^{m_{i}-1}|k_{s_{i}}\rangle $ are the eigenstates of the $i$%
-th energy level of $H_{P}$, the Hamiltonian $H$ in matrix form is: $H_{00}=%
\frac{1}{2}\omega +\varepsilon _{0}$, $H_{0i}=H_{i0}=c\sqrt{m_{i}/N}$, and $%
H_{ii}=-\frac{1}{2}\omega +E_{i}$, for $i\geq 1$; $H_{ij}=0$ for $i,j\geq 1$
and $i\neq j$.

Let the entire system evolve for time $t$, the probe qubit will exhibit a
resonance dynamics when $H_{00}=H_{11}$, or $E_{1}-\varepsilon _{0}=\omega $%
, which means the transition frequency between the reference state and the
state $|1\rangle |\varphi _{1}\rangle $ of the register $Q$ matches the
frequency of the probe qubit. Therefore by appropriately setting the
parameters $\varepsilon _{0}$ and $\omega $ and detecting the dynamics of
the probe, one can determine if the ground state energy $E_{1}$ is zero or
not, thus tell whether or not the integer $n$ is the Ramsey number. The
detailed procedures are as follows.

We start with an integer $n<R(x,y)$ and construct the corresponding problem
Hamiltonian $H_{P}$ for the $n$-vertex graph. Prepare the probe qubit in its
excited state $|1\rangle $ and the register $Q$ in the reference state $%
|\Phi \rangle $, set $\omega $ $=1$ and $\varepsilon _{0}=-1$. The entire
system of the $(L+2)$ qubits is therefore in state $|\Psi _{0}\rangle
=|1\rangle |\Phi \rangle =|1\rangle |0\rangle |0\rangle ^{\otimes L}$.
Evolving the entire system with the Hamiltonian $H$ for time $t$, then
perform a measurement on the probe qubit in basis of $|1\rangle $. Run this
procedure a number of times to obtain the probability of the probe qubit
staying in its initial state. Repeat the above steps for different evolution
time to obtain the dynamics of the probe qubit. The circuit for the
algorithm is shown in Fig.~$1$.
\begin{figure}[tbp]
\includegraphics[width=0.8\columnwidth, height=0.25\columnwidth]{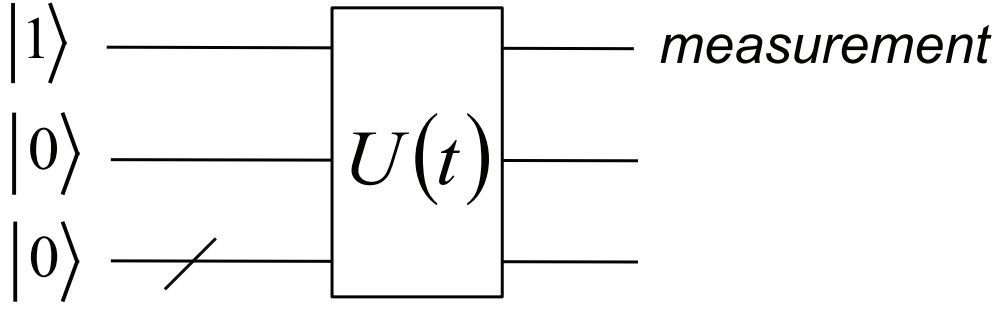}
\caption{Quantum circuit for obtaining dynamics of a probe qubit. The first
line represents the probe qubit. $U(\protect\tau )$ is a time evolution
operator driven by a Hamiltonian given in Eq.~($2$). The measurement is
perform in basis of the excited state of the probe qubit as $|1\rangle
\langle 1|$}
\end{figure}

If $n$ is not the Ramsey number, we have $E_{1}=0$ and the condition $%
E_{1}-\varepsilon _{0}=\omega $ is satisfied, the probe qubit will exhibit a
resonance dynamics. Then we can increase the integer $n$ by $1$ and repeat
the above procedures until the probe qubit shows no resonance dynamics. If $%
n $ is the Ramsey number, we have $E_{1}\geqslant 1$ and $E_{1}-\varepsilon
_{0}\geqslant 2$, there is a finite gap between the two transition
frequencies, and no resonance dynamics will be observed.

In the following, we show that the resonance and the non-resonance dynamics
of the probe qubit can be well distinguished. In the case where $n$ is not
the Ramsey number, the probe qubit will exhibit resonance dynamics. In the
algorithm, the excitation from the reference state to the ground state of $%
H_{P}$ contributes the most to the decay dynamics of the probe qubit, while
the transitions from the reference state to the excited states of $H_{P}$
also contribute to the decay of the probe qubit. The instantaneous effect of
these transitions on the decay of the probe can be mimicked by assuming that
all the states are degenerate with eigenvalue $E^{\prime }$, and $E^{\prime
} $ should be $1\leq E^{\prime }\leq v$, where $v=\max \left[ \binom{n}{x},%
\binom{n}{y}\right] $, and $\binom{n}{x}$ is the largest possible number of $%
x$-cliques in the $n$-vertex graphs, while $\binom{n}{y}$ is the largest
possible number of $y$-independent set in the $n$-vertex graphs. Suppose all
the excited states of $H_{P}$ are $\left( N-m_{1}\right) $-fold degenerate
with eigenvalue $E^{\prime }$, let $|\Psi _{2}\rangle =|0\rangle |1\rangle
\frac{1}{\sqrt{N-m_{1}}}\sum_{i=m_{1}}^{N-1}|k_{s_{i}}\rangle $, the
Hamiltonian of the algorithm $H$ in basis of $\{|\Psi _{0}\rangle =|1\rangle
|0\rangle |0\rangle ^{\otimes n},|\Psi _{1}\rangle =|0\rangle |1\rangle
|\varphi _{1}\rangle ,|\Psi _{2}\rangle \}$ can be written as
\begin{equation}
H=\left(
\begin{array}{ccc}
-\frac{1}{2} & c\sqrt{\frac{m_{1}}{N}} & c\sqrt{\frac{N-m_{1}}{N}} \\
c\sqrt{\frac{m_{1}}{N}} & -\frac{1}{2} & 0 \\
c\sqrt{\frac{N-m_{1}}{N}} & 0 & E^{\prime }-\frac{1}{2}%
\end{array}%
\right) .
\end{equation}%
The Schr\"{o}dinger equation of the above Hamiltonian can be solved exactly
for given parameters $N$, $m_{1}$, and $E^{\prime }$. In Fig.~$2$, by
setting $N=2^{10}$, $m_{1}=1$, and $c=0.02$, we show the dynamics of the
probe qubit for $E^{\prime }=1,2,5$, respectively. We can see that the probe
qubit shows clear resonance dynamics.
\begin{figure}[tbp]
\includegraphics[width=0.95\columnwidth, clip]{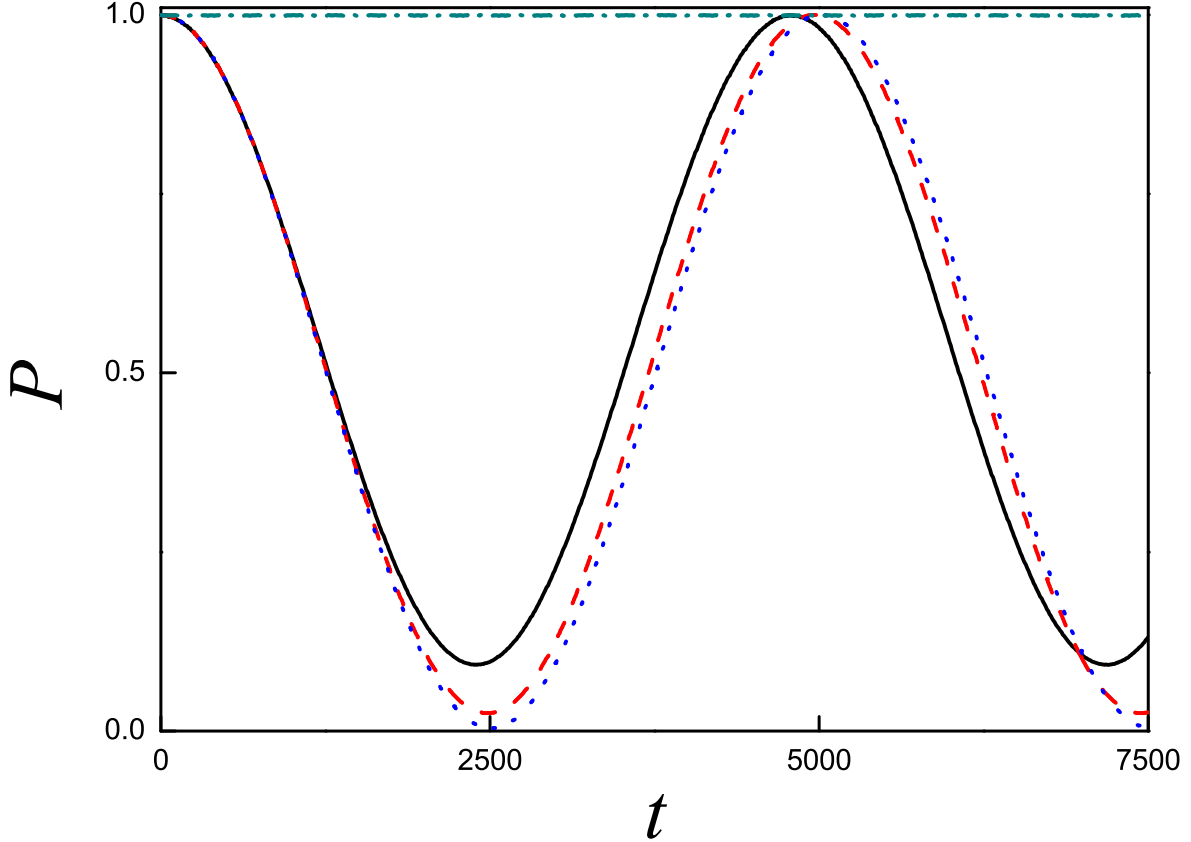}
\caption{(Color online)The probability of the probe qubit staying in its
initial excited state $|1>$ vs. the evolution time $t$, while $\protect%
\omega =1$, $\protect\varepsilon _{0}=-1$, $c=0.02$, $N=2^{10}$ and $m_{1}=1$%
. The black solid line shows the results for $E^{\prime }=1$; the red dashed
line shows the results for $E^{\prime }=2$; and the blue dotted line shows
the results for $E^{\prime }=5$, respectively. The cyan dash dot line shows
the results for the non-resonance case with $E^{\prime \prime }=1$.}
\end{figure}
\begin{figure}[tbp]
\includegraphics[width=0.9\columnwidth, clip]{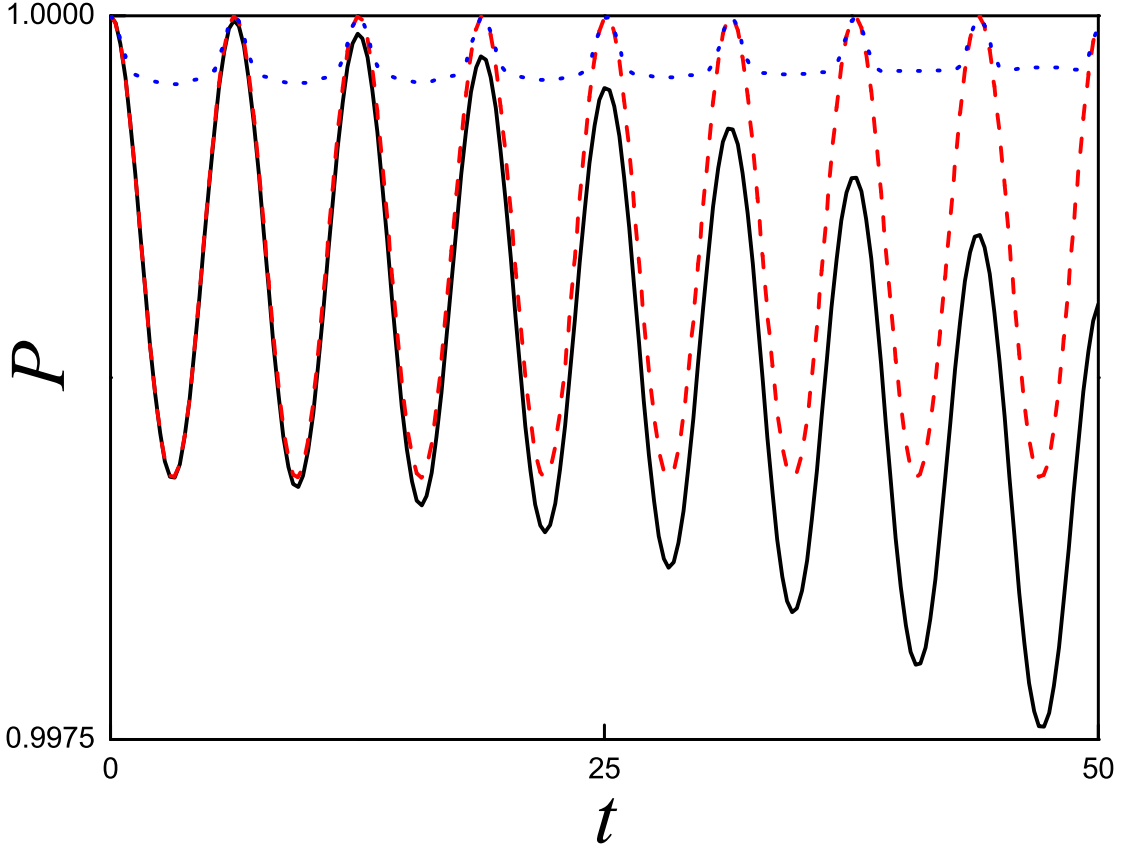}
\caption{(Color online)~Same as in Fig.~$2$. The black solid line shows the
results for the resonance case with $E^{\prime }=1$; the red dashed line
shows the results for the non-resonance case with $E^{\prime \prime }=1$;
and the blue dotted line shows the results for the case $R(2,4)=4$.}
\end{figure}

In the case where $n$ is the Ramsey number, the ground state energy of $%
H_{P} $ is $E_{1}\geq 1$. By setting $\omega $ $=1$ and $\varepsilon _{0}=-1$%
, the transition frequency between the reference state and the ground state
of $H_{P}$ is $E_{1}-\varepsilon _{0}\geqslant 2$, it does not match the
frequency of the probe qubit. Therefore no resonance dynamics will be
observed on the probe qubit. The instantaneous effect of transitions from
the reference state to all eigenstates of $H_{P}$ on the decay dynamics of
the probe qubit can be mimicked by assuming that all the eigenstates of $%
H_{P}$ are degenerate with eigenvalue $E^{\prime \prime }\geq 1$. Then the
Hamiltonian $H$ in basis of $\{|\Psi _{0}\rangle =|1\rangle |0\rangle
|0\rangle ^{\otimes n},|\Psi _{1}\rangle =|0\rangle |1\rangle \frac{1}{\sqrt{%
N}}\sum_{k=0}^{N-1}|k\rangle \}$ can be written as
\begin{equation}
H=\left(
\begin{array}{cc}
-\frac{1}{2} & c \\
c & E^{\prime \prime }-\frac{1}{2}%
\end{array}%
\right) .
\end{equation}%
With the initial state being set as $|\Psi _{0}\rangle $, the probability of
the probe qubit being in its initial state $|1\rangle $ is
\begin{equation}
P_{\text{non-res}}(t)=\frac{1}{a^{2}+4c^{2}}\left[ a^{2}+2c^{2}\left( 1+\cos
\sqrt{a^{2}+4c^{2}}t\right) \right] ,
\end{equation}%
where $a=E^{\prime \prime }-\frac{1}{2}$.

From Eq.~($6$), we can see that the minimum of $P_{\text{non-res}}(t)$ is $%
\frac{(E^{\prime \prime }-1/2)^{2}}{(E^{\prime \prime }-1/2)^{2}+4c^{2}}$.
The contribution to the decay probability of the probe qubit from the
transitions between the reference state and all the eigenstates of $H_{P}$\
must be less than $\frac{4c^{2}}{(E^{\prime \prime }-1/2)^{2}+4c^{2}}$. And
since $E^{\prime \prime }-1/2\geqslant 1/2$, this contribution can be
controlled to be very small by setting a small coupling coefficient $c$.

In Fig.~$2$, we show the dynamics of the probe qubit by setting $E^{\prime
\prime }=1$, which is the lowest possible eigenenergy of $H_{P}$ in the
non-resonance case, therefore is the largest possible contribution to the
decay of the probe qubit through transitions from the reference state to all
the eigenstates of $H_{P}$. It provides an upper-bound for the effect of the
transitions on the decay dynamics of the probe qubit. Even in this case, we
can see that the system almost stay in its initial state during the time
evolution. It can be well distinguished from the resonance dynamics of the
probe qubit as shown in the figure.

The resonance and non-resonance dynamics of the probe qubit can be
distinguished even in very short time. In Fig.~$3$, we show the dynamics of
the probe qubit for both the resonance and the non-resonance cases in very
short evolution time with the same parameters as in Fig.~$2$. We also show
the dynamics of the probe qubit for the case $R(2,4)=4$. From the figure we
can see that these three cases can be well distinguished. In the resonance
case, the probability of the probe qubit being in its excited state
decreases globally while keeping small oscillations. We can also see that
the decay probability of the probe qubit in the case $R(2,4)=4$ is much
smaller than that of the case described by Eq.~($6$).

The dynamics of the probe qubit in the resonance and the non-resonance cases
can also be distinguished through their slope in very short evolution time.
On average, the slope of the dynamics of the probe qubit is zero in the
non-resonance case, while for the resonance case, the probability of the
probe staying in its initial state is $P_{\text{res}}(t)=\cos ^{2}\left( c%
\sqrt{m_{1}/N}t\right) $ by applying perturbation theory, and the slope is
negative in short time. This makes a distinction between the resonance and
the non-resonance cases, therefore tells whether the integer $n$ is the
Ramsey number or not. The evolution time that is required to distinguish the
resonance and the non-resonance dynamics of the probe qubit is $t\sim 1/(c%
\sqrt{m_{1}/N})$, which means the runtime of the algorithm scales as $O\left(%
\sqrt{N}\right)$.

We can also find all the graphs that have minimum number of $x$-cliques or $y
$-independent sets by obtaining the ground state of the problem Hamiltonian
through the resonance dynamics of the probe qubit. We first obtain the
ground state energy of $H_{P}$ by increasing either the frequency $\omega $
of the probe qubit or the eigenvalue of the reference state $\varepsilon _{0}
$ by one each time, then run the algorithm. When the resonance dynamics on
the probe qubit is observed, this indicates that the resonance condition $%
E_{1}-\varepsilon _{0}=\omega $ is satisfied, therefore the ground state
energy $E_{1}$ of $H_{P}$ is obtained. Then we set $\omega =1$ and $%
\varepsilon _{0}=E_{1}-\omega $, the system evolves from the initial state $%
|\Psi _{0}\rangle $\ to the state $|\Psi _{1}\rangle =|0\rangle |1\rangle
|\varphi _{1}\rangle $ reaches maximal probability at time $t\sim \frac{\pi
}{2}\times 1/\left( c|\langle \Psi _{1}|\sigma _{x}\otimes A|\Psi
_{0}\rangle |\right) =\frac{\pi }{2}\frac{1}{c\sqrt{m_{1}/N}}$. The ground
state $|\varphi _{1}\rangle =\frac{1}{\sqrt{m_{1}}}%
\sum_{s_{1}=0}^{m_{1}-1}|k_{s_{1}}\rangle $ of the problem Hamiltonian is
obtained on the last $L$ qubits of the register $Q$, it is consisted of all
the states $|k_{s_{1}}\rangle $, each of which corresponds to a graph that
has minimum number of $x$-cliques or $y$-independent sets, and these basis
states can be obtained with equal probability.

The time evolution operator $U(\tau )=\exp \left( -iH\tau \right) $ can be
implemented through the Trotter formula~\cite{nc}, $U(\tau )\!=\!\left[
e^{-i\left( \frac{1}{2}\omega \sigma _{z}+H_{Q}\right) \tau /M}e^{-i\left(
c\sigma _{x}\otimes A\right) \tau /M}\right] ^{M}\!+\!O\!\left( \frac{1}{M}%
\right) $, where $M$ is a large number. The operator $e^{-i\left( \frac{1}{2}%
\omega \sigma _{z}+H_{Q}\right) \tau /M}$ is diagonal and can be implemented
efficiently. For the unitary operator $e^{-i\left( c\sigma _{x}\otimes
A\right) \tau /M}$, the operator $A=\sigma _{x}\otimes H_{d}^{\otimes L}$
can be written as $A=(I_{2}\otimes S^{\otimes n})\sigma _{x}^{\otimes
(n+1)}(I_{2}\otimes S^{\dagger \otimes n})$, where $H_{d}=S\sigma
_{x}S^{\dagger }$ and $S$ is an unitary operator. The time slices of the
unitary operator $e^{-i\left( c\sigma _{x}\otimes A\right) \tau /M}$
therefore can be written as $e^{-i\left( c\sigma _{x}\otimes A\right) \tau
/M}=(I_{2}^{\otimes 2}\otimes S^{\otimes L})e^{-ic\tau /M\sigma
_{x}^{\otimes (L+2)}}(I_{2}^{\otimes 2}\otimes S^{\dagger \otimes L})$. The
time evolution operator $e^{-ic\tau /M\sigma _{x}^{\otimes (L+2)}}$ involves
many-body interaction and can be simulated efficiently by a Hamiltonian with
two-body interactions~\cite{bravyi}.

\emph{Discussion.}--The problem of computing a Ramsey number $R(x,y)$ for
given $x$ and $y$ can be transformed to a decision problem: to determine if
the ground state energy of the problem Hamiltonian $H_{P}$ is zero or not.
In the AQO algorithm, the ground state of the problem Hamiltonian has to be
obtained and readout in order to calculate the ground state energy, this
requires extra time and resource. In our algorithm, we can conclude whether
the ground state energy of $H_{P}$ is zero or not without knowing the ground
state of $H_{P}$, thus is more convinient. And the problem of degenerate
ground state that occurs in the AQO algorithm is avoided.

This algorithm is based on a resonance phenomena and the decision problem is
solved by determining whether the probe qubit exhibits a resonance dynamics.
By appropriately setting the parameters $\varepsilon _{0}$ and $\omega $ in
the algorithm, the probe qubit will exhibit resonance dynamics if the ground
state of $H_{P}$ is zero and the integer $n$ for the corresponding $H_{P}$
is not the Ramsey number; the probe qubit will exhibit non-resonance
dynamics if the ground state energy of $H_{P}$ is not zero, which means the
corresponding integer $n\geqslant R(x,y)$. Therefore begin with an integer $%
n<R(x,y)$, one can find the Ramsey number $R(x,y)$ for given $x$ and $y$
from the dynamics of the probe qubit.

In this algorithm, the probe qubit can detect the energy spectrum of the
problem Hamiltonian $H_{P}$ by varying the probe frequency or the reference
state energy. The dynamics of the probe is the "fingerprint" of the problem
Hamiltonian. Especially for $H_{P}$ that has discrete energy spectrum, the
dynamics of the probe qubit are very different in the resonance and the
non-resonance cases and can be well distinguished in time scale of $O\left(
\sqrt{N}\right) $. Based on this, the ground state energy and the ground
state of the problem Hamiltonian can also be obtained. And since we consider
all possible excitations from the reference state to the basis states of the
problem Hamiltonian with equal probability, the ground state we obtained is
an unified superposition of the basis states that represent the graphs
having minimum number of $x$-cliques and $y$-independent sets. This
algorithm can also be applied for solving some other decision problems, such
as the $3$-SAT problems.

\begin{acknowledgements}
This work was supported by \textquotedblleft the Fundamental Research Funds for the Central Universities\textquotedblright\ of China and the National Nature Science Foundation of China~(Grants No.~11275145 and No.~11305120).
\end{acknowledgements}

\end{document}